\documentclass[12pt]{article}
\usepackage {graphicx}

\begin{document}

\begin{center}
\textbf{Twin Boundaries and Heat Capacity of a Crystal}

\bigskip
V.N.Dumachev \footnote{e-mail: dumv@comch.ru}

\bigskip
\footnotesize{Voronezh Institute of the Ministry of Internal
Affairs of the Russian Federation, Voronezh, Russia }
\end{center}

\bigskip

\begin{abstract}
The influence of twin boundaries on the heat capacity and diffuse
scattering in a crystal is described within the framework of a
macroscopic dynamic theory. \copyright  2005 Pleiades Publishing,
Inc.
\end{abstract}

\bigskip
Achievements of the modern nanotechnologies in the synthesis of
structures of arbitrary dimensions set the important task of
studying natural low-dimensional formations in crystals, such as
dislocations and domain walls, and the influence of such objects
on the mechanical and kinetic properties of materials. This paper
shows that coherent twin boundaries can significantly influence
the thermodynamic properties of crystals.

Consider an elastic continuous medium containing a defect that
acts as the source of a plastic deformation field. A local change
in position of the surface of this defect will modify the mismatch
between the defect and environment. In the case of a planar
defect, this is manifested by the motion of misfit dislocations.
The resulting configurational or surface forces act so as to bring
the defect boundary to an equilibrium state, while the mismatch
becomes (due to inertial properties of the medium) a source of
elastic waves. Considering the total deformation of the crystal
$u_{lm} $ in the presence of the structural defect as comprising
the elastic $\varepsilon _{lm} $ and plastic $s_{lm} $ components,
the Lagrangian of the elastic continuum can be written as

\begin{equation}
\label{eq1} L = \frac{{1}}{{2}}\int {d\mathbf{r}dt\left( {\rho
\left( {\dot {u}_{i}} \right)^{2} - \left( {u_{ik} - s_{ik}}
\right)\lambda _{iklm} \left( {u_{lm} - s_{lm}} \right)} \right)}
.
\end{equation}

Let us consider the coherent boundary as an independent object in
the crystal possessing its own dynamical variable $\varsigma
\left( {\mathbf{r}_{||} ,t} \right)$ representing a deviation of
the boundary relative to the habit plane. Then, writing the
plastic deformation caused by this deviation as

\[
s_{ij} = \varsigma \left( {\mathbf{r}_{||} ,t} \right)\delta
\left( {z} \right)\left[ {S_{ij}}  \right]
\]

\noindent substituting this expression into Lagrangian
(\ref{eq1}), and considering variations over the dynamical
variables $u_{i} \left( {\mathbf{r},t} \right)$ and $\varsigma
\left( {\mathbf{r}_{||} ,t} \right)$, we eventually obtain a
system of differential equations [1]

\[
\begin{array}{l}
 \rho \ddot {u}_{i} - \lambda _{iklm} \partial _{kl}^{2} u_{m} + \partial
_{k} \left( {\lambda _{ik}^{s} \delta \left( {z} \right)\varsigma
\left(
{\mathbf{r}_{||} ,t} \right)} \right) = 0 \\
 \left( {\lambda _{ik}^{s} \partial _{k} u_{m} - \lambda ^{s}\delta \left(
{z} \right)\varsigma \left( {\mathbf{r}_{||} ,t} \right)} \right)_{z = 0} = 0 \\
 \end{array}.
\]

\noindent These equations have a self-consistent solution under
the condition provided by the dispersion relation [2]

\begin{equation}
\label{eq2} \frac{{q_{y}^{2} - \omega ^{2}/c_{t}^{2}} }{{\sqrt
{\mathbf{q}_{||}^{2} - \omega ^{2}/c_{t}^{2}} } } +
\frac{{4q_{x}^{2}} }{{\omega ^{2}/c_{t}^{2}} }\left( {\sqrt
{\mathbf{q}_{||}^{2} - \omega ^{2}/c_{l}^{2}}  - \sqrt
{\mathbf{q}_{||}^{2} - \omega ^{2}/c_{t}^{2}} } \right) = 0 .
\end{equation}

Here, it is assumed that the habit plane of the boundary is
perpendicular to the $0z$ axis, $\left[ {S_{ik}}  \right] =
1/2\left( {n_{i} S_{k} + n_{k} S_{i}} \right)$ is the offset of
the plastic deformation tensor on the pas- sage through the
boundary, $n_{i} = \left( {0,0,1} \right)$ is vector of the normal
to the boundary, $S_{i} = \left( {S_{x} ,0,0} \right)$ is the
vector of the boundary displacement, and $\lambda ^{s} = S_{ik}
\lambda _{iklm} S_{lm} = S_{ik} \lambda _{ik}^{s} $. The roots of
dispersion equation (\ref{eq2}) determine a relation between the
eigenvector $\mathbf{q}_{||} $ and the intrinsic frequency $\omega
$ of bending oscillations of the boundary [3]: $\omega = \xi
\left( {\varphi}  \right)c_{t} \mathbf{q}_{||} $ (fig.1).

\begin{figure}[tbp]\centering
  \includegraphics[]{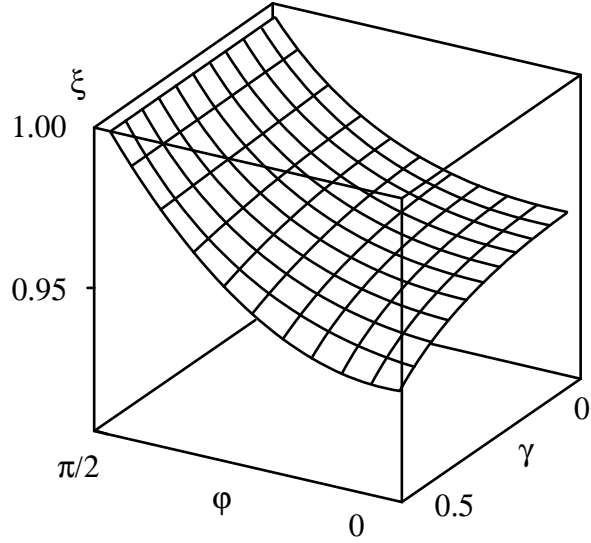}
  \caption{Dependence of the $  \xi =\omega /(
c_{t} \mathbf{q}_{||}) $ on angle $\varphi=Atan(S_{y}/S_{x})$ and
$\gamma=c_{t}/c_{l}$.}
\end{figure}

First, let us calculate the phonon heat capacity of a crystal [4].
The Einstein approximation for this prob lem stipulates a linear
dependence of the wave fre quency $\omega $ on the wave vector
$\mathbf{q}$. Using Eq. (\ref{eq2}) it is also possible to take
into account the influence of a nonlin ear dispersion related to
the presence of a planar defect in the crystal studied. The number
of field oscillators with the wavenumbers below $\mathbf{q}$ can
be determined as

\[
N\left( {\omega}  \right) = \int\limits_{0}^{\mathbf{q}}
{d\mathbf{q}_{||}}  = \int\limits_{0}^{\omega /c_{t} \xi \left(
{\varphi}  \right)} {qdqd\varphi} = \frac{{\omega
^{2}}}{{2c_{t}^{2}} }\int\limits_{0}^{\pi /2} {\frac{{d\varphi}
}{{\xi ^{2}\left( {\varphi}  \right)}}} = \frac{{\omega
^{2}}}{{2c_{t}^{2}} }\frac{{\pi} }{{2}}\Omega
\]

\noindent where $\Omega \cong 1.074$. Then, the Debye frequency
$\omega _{0} $ for the bound phonon states is

\[
\omega _{0}^{2} = \frac{{4Nc_{t}^{2}} }{{\Omega \pi} } =
\frac{{\omega _{\theta} ^{2}} }{{3\Omega} },
\]

\noindent where $\omega _{\theta}  $ is the Debye frequency of the
bulk waves. Such a considerable decrease in the Debye frequency of
the twin boundary is related to softening of the phonon modes
accompanying the displacement of atoms in the crystal lattice
along the direction of the shear vector $S_{i} $.

Using the dispersion relation (\ref{eq2}), we obtain the average
energy of localized oscillations at low temperatures $\left( {T <
< T_{0}}  \right)$ in the Debye approximation:

\[
\left\langle {E_{0}}  \right\rangle = \int\limits_{0}^{\infty}
{\frac{{\hbar \omega} }{{exp\left( {{\raise0.7ex\hbox{${\hbar
\omega} $} \!\mathord{\left/ {\vphantom {{\hbar \omega}
{kT}}}\right.\kern-\nulldelimiterspace}\!\lower0.7ex\hbox{${kT}$}}}
\right) - 1}}dN\left( {\omega}  \right)} = \frac{{\hbar}
}{{c_{t}^{2}} }\frac{{\pi }}{{2}}\Omega \omega _{0}^{3} 2\zeta
\left( {3} \right) = \frac{{1}}{{\sqrt {3\Omega} } }\left\langle
{E_{\theta} }  \right\rangle ,
\]

\noindent where $\zeta \left( {3} \right) = 1.202057$ is the
Riemann zeta function and $\left\langle {E_{\theta} }
\right\rangle $ is the average energy of the bulk waves in the
defect-free crystal.

The heat capacity is defined as $c_{v}^{0} = \partial E/\partial
T$. However, in our case, it is convenient to calculate the ratio
of this quantity to the heat capacity of the defect-free crystal:

\begin{center}
 $\frac{{c_{v}^{0}} }{{c_{v}^{\theta} } } = \frac{{\partial E_{0} /\partial
T}}{{\partial E_{\theta}  /\partial T}} = \frac{{\partial E_{0}}
}{{\partial E_{\theta} } } = \frac{{1}}{{\sqrt {3\Omega} } }$  or
$c_{v}^{0} = \frac{{c_{v}^{\theta} } }{{\sqrt {3\Omega} } }$.
\end{center}

Evidently, a decrease in the heat capacity is related to two
factors: (i) a threefold decrease in the number of bulk
oscillatory modes of the crystal and (ii) the anisotropic velocity
of the surface wave packet propagating along the twin boundary.

The presence of a single coherent twin boundary in the crystal
does not influence the positions of electron or X-ray diffraction
reflections.An ensemble of parallel twins will only lead to the
appearance of superstructure satellites. However, a change in the
spectrum of crystal oscillations in the vicinity of the twin
boundary leads to a significant decrease in the total intensity of
the Born scattering as a result of an increase in the diffuse
scattering component [5].

Considering the thermal oscillations of atoms at the twin boundary
as a system of elastic standing waves obeying relation
(\ref{eq2}), the influence of these oscillations on the intensity
of selective maxima can be described in terms of the Debye–Waller
factor

\[
2M = \frac{{16\pi ^{2}\left\langle {u^{2}} \right\rangle
}}{{3}}\frac{{sin^{2}\vartheta} }{{\lambda ^{2}}}.
\]

Here, the mean square deviation of atoms in the crystal can be
expressed via the above relation for the average energy as

\[
\left\langle {u_{0}^{2}}  \right\rangle = \int\limits_{0}^{\infty}
{\frac{{E_{0}} }{{\omega ^{2}}}dN\left( {\omega}  \right)} =
\int\limits_{0}^{\infty}  {\frac{{\hbar} }{{exp\left(
{{\raise0.7ex\hbox{${\hbar \omega} $} \!\mathord{\left/ {\vphantom
{{\hbar \omega}
{kT}}}\right.\kern-\nulldelimiterspace}\!\lower0.7ex\hbox{${kT}$}}}
\right) - 1}}} \frac{{dN\left( {\omega}  \right)}}{{\omega} } =
\sqrt {3\Omega}  \left\langle {u_{\theta} ^{2}}  \right\rangle .
\]

Similar to the case considered above, the presence of a coherent
twin boundary in the crystal changes the observed effect by a
factor of $\sqrt {3\Omega}  \approx 1.8$.

\bigskip
\textbf{References}

1. A. M. Roshchupkin, V. N. Nechaev, and V. N. Dumachev, Izv.
Ross. Akad. Nauk, Ser. Fiz. 59 (10), 108 (1995).

2. V. N. Nechaev and A. M. Roshchupkin, Sov. Phys. Solid State
31,1321 (1989).

3. L. D. Landau and E. M. Lifshitz, Course of Theoretical Physics,
Vol. 7: Theory of Elasticity (Pergamon, New York, 1986).

4. Ya. P. Terletskii, Statistical Physics, (3rd Edition)
(North-Holland, Amsterdam, 1971).

5. V. I. Iveronova and G. P. Revkevich, Theory of X-Ray Scattering
(Moscow State University, Moscow, 1978) [in Russian].

\bigskip
\footnotesize{Translated by P. Pozdeev}

\end{document}